\documentclass[fdp,a4paper,fleqn%
]{w-art}
\usepackage{times,cite}
\usepackage[]{graphicx}
\usepackage{epic,pst-plot,amssymb}        \def\mao#1{\mathop{\rm #1}\nolimits} 
\def\BC{\begin{center}}      \def\EC{\end{center}}
\def\BE {\begin{equation}}      \def\EE {\end{equation}}
\def\BEA{\begin{eqnarray}}      \def\EEA{\end{eqnarray}}
\def\BI{\begin{itemize}}     \def\EI{\end{itemize}}   \def\bye{\end{document}}
\def\BP{\begin{picture}}     \def\EP{\end{picture}}    \let\lila=\magenta
\def\putc#1)#2{\put#1){\makebox(0,0)[c]{#2}}}            
\def\putm#1)#2{\put#1){\makebox(0,0)[c]{$#2$}}}
\def\Pput#1)#2{\BP(0,0)\put#1){#2}\EP}          \long\def\del#1\enddel{}

\unitlength=1pt \psset{unit=1pt}        

\def\putvec#1,#2,#3,#4,#5){\put#1,#2){\vector(#3,#4){#5}}}
\def\putlab#1)#2#3{\put#1){\makebox(0,0)[#2]{\small #3}}}
\def\VR#1#2{\vrule height #1mm depth #2mm width 0pt}
\def\TVR#1#2{@{~~\VR{#1}{#2}}}

\def\VS#1 {\vspace*{#1pt}}   \def\HS#1 {\hspace*{#1pt}}  

\def\IZ{{\mathbb Z}}    \def\IC{{\mathbb C}}    \def\IR{{\mathbb R}}
\def\IP{{\mathbb P}}	\def\IF{{\mathbb F}}	\def\IT{{\mathbb T}}

\def\co{{\cal O}}  \def\cu{{\cal U}}

\let\bra=\langle   \let\ket=\rangle     \let\then=\Rightarrow   
\def\2{{1\over2}}  \let\5=\bar  \let\6=\partial \def\7#1{{#1}\llap{/}}  

\let\a=\alpha   \let\b=\beta          %% --> GREEK
    \let\h=\eta     \let\th=\theta  \let\e=\varepsilon
   \let\l=\lambda  \let\m=\mu      
            \let\p=\pi      \let\r=\rho     \let\s=\sigma 
               
               \let\S=\Sigma 
             \let\D=\Delta

\def\ifundefined#1{\expandafter\ifx\csname#1\endcsname\relax}

     \let\fns=\footnotesize	\let\ex=\times	\let\N=\nabla
	\def\qtq#1 {\qquad\hbox{#1}\qquad}

\begin{document}
%%    The information for the title page will be placed between
%%    \begin{document} and \maketitle. The order of most entries
%%    is determined by the class file and can not be changed by
%%    rearranging them. The maketitle command follows after the
%%    abstract.
%%
%%    Most of the following commands will be completed by the publisher.
%%
%%    The copyrightyear is defined in the .clo file as the first argument
%%    of the copyrightinfo command. If the copyrightyear differs from that
%%    value it might be adjusted by the following definition:
%%
%% \renewcommand{\copyrightyear}{2007}% uncomment to change the copyrightyear.
%%
\DOIsuffix{theDOIsuffix}
%%
%% issueinfo for the header line
\Volume{55}
\Month{01}
\Year{2007}
%%
%%    First and last pagenumber of the article. If the option
%%    'autolastpage' is set (default) the second argument may be left empty.
\pagespan{1}{}
%%
%%    Dates will be filled in by the publisher. The 'reviseddate' and
%%    'dateposted' (Published online) entry may be left empty.
\Receiveddate{XXXX}
\Reviseddate{XXXX}
\Accepteddate{XXXX}
\Dateposted{XXXX}
\keywords{Conifold Transitions, Mirror Symmetry, Calabi-Yau, Toric Geometry.}
\subjclass[pacs]{02.10.De 02.40.-k 02.40.Tt 11.25.-w 11.25.Mj 12.60.-i}

%% \pretitle{Editor's Choice}

%% We have a short and a long form for the title. The short form
%% (optional argument) goes into the running head.

\title[Beyond toric hypersurfaces]{The making of Calabi-Yau spaces: 
					Beyond toric hypersurfaces}

\author[M. Kreuzer]{Maximilian Kreuzer\inst{1,}%
  \footnote{Corresponding author\quad 
		E-mail:~\textsf{maximilian.kreuzer@tuwien.ac.at},
            Phone: +43\,1\,58801\,13621,
            Fax: +43\,1\,58801\,13699}}
\address[\inst{1}]{Institute for Theoretical Physics, TU--Wien, 
        Wiedner Hauptstr. 8--10, 1040 Vienna, AUSTRIA}
%%    \dedicatory{This is a dedicatory.}
\dedicatory{					% {\it Acknowledgements.}
	This work is supported in part by the {\it Austrian Research Funds} 
	FWF under grant Nr. P18679.
}

\begin{abstract}
	While Calabi-Yau hypersurfaces in toric ambient spaces provide a
	huge number of examples, theoretical considerations as well as
	applications to string phenomenology often suggest a broader 
	perspective.
	With even the question of finiteness of diffeomorphism types
	of CY 3-folds unsettled, an important idea is Reid's conjecture that 
	the moduli spaces are connected by certain singular transitions.
	We summarize the results of our recent construction of a large class
	of new CY spaces with small Picard numbers and of their mirrors via 
	conifold transitions and discuss the benefits of other approaches 
	to interesting locations in the web that have been or should be 
	pursued.
\end{abstract}
%% maketitle must follow the abstract.
\maketitle                   % Produces the title.

\section{Introduction}

Examples of Calabi-Yau manifolds are of obvious interest in mathematics
and in physics. A first sizable set of almost 8000 was constructed by  
enumeration of complete intersections in products of projective 
spaces \cite{CICY}. Mirror symmetry of the Hodge data, however, 
% i.e. the exchange of $h_{11}$ and $h_{12}$, 
only emerged for hypersurfaces of {\it weighted} projective spaces 
\cite{CLS}, where the resolution of ambient space singularities contributes
positively to the Euler number. While neither the complete list of transversal
weights \cite{Kreuzer:1992da,Klemm:1992bx} nor orbifolding 
\cite{Kreuzer:1992iq} provided the 				%still 
missing mirror manifolds, the proper generalization
was discovered by Batyrev \cite{Bat}, who found a combinatorial criterion for 
hypersurfaces in toric varieties to have trivial canonical class. %bundle.
Mirror symmetry, which allows to compute instanton corrections in
physics, or quantum cohomology in mathematics \cite{Can,CK,Hori},
thus corresponds to the exchange of
a dual pair of reflexive lattice polytopes \cite{Bat}. A generalization 
of this combinatorial incarnation of mirror symmetry was found by Borisov
\cite{bo93} and proven at the level of Hodge numbers for the corresponding
complete intersections in toric ambient spaces by Batyrev 
and~Borisov~\cite{BBnef,BBsth,BBgen,BN}.

Not much is known about how a complete classification of Calabi-Yau manifolds
could be achieved and even finiteness of the number of moduli spaces is an
open question. A theorem by Wall \cite{Wall} states that the Hodge data
together with the tripple intersections and the second Chern class completely
determine the diffeomorphism type of a simply connected Calabi-Yau 3-fold.
%\footnote{~
	(In all known examples instanton numbers, which depend on the 
	symplectic structure, agree for diffeomorphic CYs \cite{kkrs,ccy}.) 
%}

According to a conjecture of Miles Reid \cite{Reid} the moduli spaces of all
Calabi-Yau 3-folds are connected by singular transitions. 
A possible strategy for exploring new realms in this web would hence be
to start from a large class of examples and to study singular transitions.
Our starting point along these lines is the list of hypersurfaces in
toric varieties whose fan is determined by a reflexive polytope.
These polytopes have been enumerated in up to 4 dimensions 
\cite{crp,c3d,c4d,pwf}, where the total number is 473\,800\,776. Starting
from this huge list we look for conifold singularities that are 
combinatorially encoded
% This is the case for curves of conifolds, which correspond to 
% 2-faces that are parallelograms $\th^\circ$ of minimal volume. 
% These singular curves generically intersect with the Calabi-Yau in distinct 
% conifold points whose number is determined by the length of the 
% dual edge $\th$. 
%Taking into account the smoothing condition we 
and	find 30241
new cases of Calabi-Yau varieties with small Picard numbers, 
% $h_{11}\sim3$,
including 68 distinct topologies for $h_{11}=1$; % \cite{ccy}. 
we also propose a mirror construction and compute the Picard-Fuchs 
operators in many cases \cite{ccy}.

In the present note we summarize the result of our construction of new
Calabi-Yau manifolds via conifold transitions and comment on other recent 
work. In section 2 we fix our notation by explaining the combinatorics of 
toric geometry and discuss singularities and the conifold transition.
In section 3 we review our results on conifold Calabi--Yau's.
In section 4 we conclude by discussing different approaches and some of 
their benefits as well as open problems and ideas that would be interesting 
to pursue.

\section{Geometry and combinatorics}

One of the most studied Calabi-Yau manifolds is the quintic, i.e. the 
hypersurface in $\IP^4$ that is defined by the vanishing of a
homogeneous polynomial $p_d(z_j)=0$ of degree $d=5$. We denote the 
homogeneous coordinates of projective points by 
$[z]=(z_0:z_1:z_2:z_3:z_4)\in\IP^4$, where we use square brackets for 
equivalence classes w.r.t. the 
$\IC^*$-identification $(z_j)\to(\l z_j)$ for $\l\in\IC^*=\IC\setminus\{0\}$.
$\IP^n$ is covered by affine patches $\cu_i=\IP^n\setminus D_i$ with
$D_i=\{[z]\in\IP^n : z_i=0\}$. On $\cu_i$ we can set $z_i=1$ and
identify the remaining coordinates with affine coordinates on 
$\IR^n\equiv U_i\subset\IP^n$. The vanishing of a homogeneous coordinate, 
like the vanishing of any homogeneous polynomial, is independent of the
representative. On each patch $\cu_i$ the hypersurface $p(z)=0$ is the 
vanishing set of the {\it function} $f_i([z])=p(z)/(z_i)^d$; hence $p(z)$
is a section of the line bundle $\co(d)$ with transition functions 
$g_{ij}=(z_i/z_j)^d$.

Introducing affine coordinates $t_i=z_i/z_0$ for $U_0$ we
observe that all patches contain the points with all $t_i\neq0$, i.e.
$(t)\in(\IC^*)^4$. The multiplicative
group $(\IC^*)^n$ is called the (algebraic) $n$-torus $\IT^n$ because it is 
the complexification of $(U(1))^n$. Projective space
$\IP^4=\IT^4\cup D_0\cup\ldots\cup D_4$
is hence $\IT^4$ plus limit points, which are obtained by the vanishing 
of homogeneous coordinates.
This generalizes to toric varieties $\IP_\S$, which are (partial) 
compactifications of $\IT^n$ with the natural torus action on itself
extending to~$\IP_\S$ (cf. \cite{Kreuzer:2006ax,Closset:2009sv,CK,Hori} and
references therein). The points of {\it weighted} projective spaces 
$W\IP^n_{q_0\ldots q_n}$, i.e. $\IC^*$-equivalence classes w.r.t. weighted
scalings $(z_j)\to (\l^{q_j}z_j)$, are of this form. General toric varieties 
are then a further generalization with $r$ homogeneous coordinates and 
$r-n$ weighted scalings.

\subsection{Polytopes and homogeneous coordinates}

The data of a toric variety $\IP_\S$ of dimension $n$ is given by a fan 
$\S$, which is a collection of cones $\s\in\S$ that is closed under 
intersections and taking faces. Moreover, all cones 
$\s\subseteq N_\IR\cong\IR^n$ have to be rational 
(i.e. generated by lattice vectors in $N\cong\IZ^n$) and 
strongly convex (i.e. must not contain lines / have an apex). 
We now show how this data relates to the homogeneous coordinates.

Laurent polynomials $f(t)=\sum_{{m}\in\IZ^n} ~c_m~t^m$ are finite
sums of Laurent monomials $t^m=t_1^{m_1}\ldots t_n^{m_n}$ with exponent 
vectors $m\in M\cong\IZ^n$. 
The Newton polytope $\D_f$ of a Laurent polynomial $f$ is the convex hull
of the exponent vectors $m\in M$ in the real extension 
$\IR^n\cong M_\IR\supseteq M$ of the lattice $M$. Hence 	\VS-6
\BE
	f(t)=\sum_{m\in\D_f\cap M}~c_m\,{t}^{m}\qquad\hbox{with}
	\quad t^m=t_1^{m_1}\ldots t_n^{m_n},\quad m\in M\cong\IZ^n.
\EE								\\[-12pt]
For a toric variety the affine coordinates $t_i$ are Laurent monomials in the
homogeneous coordinates $z_j$,						\VS-8
\BE	\!\!\!\!\!\!\!\!					\label{tz}
	t_i=\prod_{j=1}^r z_j^{v_{ji}},\quad v_{ji}=\bra v_j,e_i\ket\in\IZ
	~~\quad \then	~~\quad	t^m=\prod _{j=1}^r z_j^{\bra v_j,m\ket},
	~~\quad v_j\in N=\mao{Hom}(M,\IZ),
	%%%	\quad\nabla=\bra v_1\ldots v_r\ket_{conv},
\EE								\\[-12pt]
so that the homogeneous coordinates $z_j$ are canonically associated
to exponent vectors $v_{j}$ with components $\bra v_j,e_i\ket=v_{ji}$
where $\{e_i\}$ is the canonical basis of $M\cong\IZ^n$. The $v_j$ hence
naturally live in the lattice $N$ dual to $M$. The vectors $v_j\in N$ are 
the primitive generators of the 1-dimensional cones $\r_j\in\S(1)\subseteq\S$,
where the $k$-skeleton $\S(k)$ consists of the $k$-dimensional cones of $\S$.
Weighted scalings				\label{q}	\VS-7
\BE	\textstyle			
	(z_1,\ldots,z_r)\to (\l^{q_1	%^{(a)}
	}z_1,\ldots,\l^{q_r		%^{(a)}
	}z_r)\qtq with \sum\limits_{j\le r}\;q_j %^{(a)}
	v_j=0\in N				
\EE								\\[-11pt]
of the homogeneous coordinates $z_j$ leave the affine
coordinates $t_i$ (and hence the points of the dense subset 
$\IT^n\subset\IP_\S$) invarint if and only if $\sum q_j v_j=0$. 
Maximal linearly independent sets of integral solutions 
$q_j^{(a)}$, $a=1,\ldots,r-n$ to this equation are called charge vectors
in physics because they correspond to charges of chiral superfields in
the gauged linear sigma model (GLSM) \cite{wi93}. 
We thus can think of $\IP_\S$ in four different ways:	\VS-3
\BE
	\IP_\S	=(\IC^r-Z_\S)/(\IC^*)^{r-n}\ex G
		=\IT^n\cup\mao{\hbox{$\bigcup$}}\limits_{j\le r} D_j
		=\mao{\hbox{$\bigcup$}}\limits_{\s\in\S} \cu_\s
		=\IC^r/\!\!/\,U(1)^{r-n}.		
\EE								\\[-10pt]
A subtle point of the {\it holomorphic quotient} %construction 
$(\IC^r-Z_\S)/(\IC^*)^{r-n}\ex G$ is an additional
discrete identification by %an 
a finite abelian group 			
$ %\BE
	G\cong N/\mao{span}_\IZ(v_1,\ldots,v_r)
$ %\EE								\\[-15pt]
if the vectors $v_j$ do not span the $N$-lattice. Taking quotients by 
phase symmetries thus corresponds to a refinement of the $N$-lattice.
The exceptional set $Z_\S$, which has to be removed before taking the 
quotient, is the union of irreducible components 
$Z_I=D_{I_1}\cap\ldots\cap D_{I_s}$, where $I$ are minimal 
index sets such that there is no cone $\s\in\S$ containing all 
$v_{I_1},\ldots,v_{I_s}$.%
\footnote{~ 
	These minimal index sets $I$ have been called primitive collections
	by Batyrev.
}
%In simpler words, 
Succinctly,
a subset of homogeneous coordinates may vanish 
simultaneously $z_{j_1}=\ldots=z_{j_k}=0$ if and only if there exists a cone 
$\s$ 
containing all corresponding rays $\r_{j_1},\ldots,\r_{j_k}\subseteq \s$. 
For each cone $\s\in\S$ we thus define an affine patch 
$U_\s=\IP_\S\setminus\mao{\hbox{$\bigcup$}}\nolimits_{v_j\not\in\s}D_j$, i.e.
$\cu_\s$ consists of the points $[z]$ for which all coordinates $z_j$ with
$\r_j\subseteq\llap{\large$\slash$ }\s$ are nonzero $z_j\neq0$.
The ring of regular functions on $\cu_\s$ is the semigroup ring 
$A_\s=\IC(M\cap \s^\vee)$ of the lattice points in the dual cone 
  %\VS-4
$ %\BE		
	\s^\vee=\{\,x\in M_\IR~:~\bra y,x\ket\ge0\quad\forall y\in\s\,\}, 
$ %\EE							\\[-15pt]
because $t^m=\prod z_j^{\bra v_j,m\ket}$ %\ref{tz}
is regular on $\cu_\s$ if and only if $m\in\s^\vee$. 
The construction of $\cu_\s$ as
the spectrum of $A_\s$ is, in fact, more general than the holomorphic 
quotient, which literally only works for simplicial fans \cite{CoxHCR}.
The {\it symplectic} quotient $\IC^r/\!/\,U(1)^{r-n}$ is related to
the GLSM \cite{wi93} with gauge group $U(1)^{r-n}$. It modes out only the 
compact part of the $\IC^*$ identifications while the radial directions are 
fixed to regular values of the moment maps $\m_a=\sum_j q_j^{(a)}|z_j|^2$, 
which provide parameters for the K\"ahler metric
\cite{wi93,Hori,Kreuzer:2006ax,Closset:2009sv}.

\subsection{Singularities, blow-ups and the conifold transition}

Singularities of $\IP_\S$ are visibe in affine patches. Thus they are	% and 
encoded in the combinatorics of the cones $\s$: 		% as follows:
\\[2pt]
{\bf Theorem:} A toric variety $\IP_\S$ is smooth if and only if all cones are
simplicial and basic, i.e. all cones $\s\in\S(k)$ are generated by $k$ vectors
$v_{j_1},\ldots,v_{j_k}$ that are part of a lattice basis of the $N$-lattice. 
\\[2pt]
{\bf Theorem:} A toric variety $\IP_\S$ is compact if and only if the support 
of the fan $\S$ covers $N_\IR$.

In two dimensions all cones are simplicial so that, up to a change of basis,
all singularities come from a cone of the form
$\s\!=\!\IR_+\!\left({1\atop0}\right)+\IR_+\!\bigl({\!-q\atop \,p}\bigr)$
with $\mao{gcd}(p,q)=1$ and $q<p$ of volume $p$ (in lattice units).
% of basic simplices). 
The dual cone
$\s^\vee\!=\!\IR_+\!\bigl({0\atop1}\bigr)+\IR_+\!\bigl({p\atop q}\bigr)$
is generated by the exponent vectors $m$ of the regular monomials 
$t^m=z_2^p,z_1z_2^{p-q},\ldots,z_1^p$, which are invariant under 
$\IZ_p:(z_1,z_2)\to(e^{{2\p i}\frac qp}z_1,e^{{2\p i}\frac 1p}z_2)$. Hence
$\cu_\s=\IC^2/\IZ_p$.

\begin{figure}
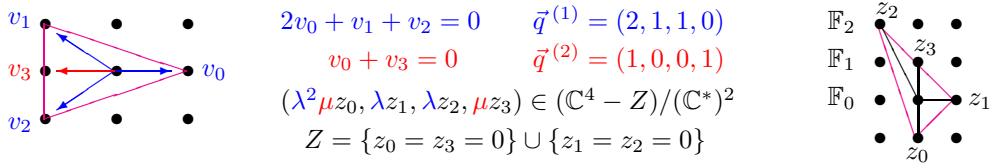
	\unitlength=1.8pt 	% \psset{unit=1.8pt}
\BC
\BP(202,28)(-20,-12)     \matrixput(-13,-10)(15,0)3(0,10)3{\circle*2} 
\lila 	\put(.3,0){\drawline(15,0)(-15,10)
		\drawline(15,0)(-15,-10)\drawline(-15,10)(-15,-10)}
\blue	\linethickness{.6pt}\putvec(0,0,1,0,12)
		\put(-22.3,-12){$v_2$}\put(18.5,-1){$v_0$}
        	\putvec(0,0,-3,-2,12)\putvec(0,0,-3,2,12)
		\red\putvec(0,0,-1,0,12)
        	\red\put(-22.3,-1){$v_3$}
		\blue\put(-22.3,10){$v_1$}

\blue   \put(35,9){$2v_0+v_1+v_2=0$}	\put(88,9){$\vec q^{~(1)}=(2,1,1,0)$}
\red    \put(45,1){$v_0+v_3=0$}		\put(88,1){$\vec q^{~(2)}=(1,0,0,1)$}
\black	\put(35,-8){$(\blue\l^2\red\m\black z_0,\blue\l\black z_1,
                \blue\l\black z_2,\red\m\black z_3)\in(\IC^4-Z)/(\IC^*)^2$}
	\put(40,-16){$Z=\{z_0=z_3=0\}\cup\{z_1=z_2=0\}$}

\put(175,-6){\put(0,0){
\lila   \drawline(0,0)(-8,-8)\drawline(0,0)(-16,16)\drawline(-8,-8)(-16,16)}
\black  \put(-25,-1){$\mathbb F_0$}\put(-25,7){$\mathbb F_1$}
        \put(-25,15){$\mathbb F_2$}
        \put(4,0){\matrixput(-18,-8)(8,0)3(0,8)4{\circle*2}
        \put(-2,0){\drawline(0,0)(-8,0)\drawline(-8,-8)(-8,8)
                \drawline(-16,16)(-8,0)}}
        \put(-8.7,-12.3){$z_0$}\put(4.5,-1){$z_1$}\put(-15,18){$z_2$}
        \put(-7,10.3){$z_3$}						}
\EP\EC	\black
\caption{The weighted projective space $W\IP^2_{211}$ and the blow-up of its 
	singular point, the Hirzebruch surface $\IF_2$.}	\label{fig:WP}
\VS-9
\end{figure}

All singularities of toric varieties can be resolved by subdividing cones, 
which is done by adding new rays and triangulating. 
We illustrate this in Fig.\,\ref{fig:WP} 
with the compact example of $W\IP_{211}^2$, whose weighted $\IC^*$
identification has the fixed point $(1:0:0)$ for $\l=-1$. Subdivision of
the cone $(v_1,v_2)$ with volume 2 by $v_3=\frac12(v_1+v_2)$ changes
the exceptional set and adds the coordinate $z_3$. For $z_3\neq0$ we can
set $z_3=1$ and recover all points of $W\IP_{211}^2$ except for its singular
point $(1:0:0:1)$, which gets replaced by its blow-up $\IP^1\ni(1:z_1:z_2:0)$,
i.e. by the points with $z_3=0$ for which we can set $z_0=1$.

\begin{figure}%[htb]
\unitlength=1pt							\del
\BP(260,60)(-52,-26)\unitlength=2pt
	\put(0,0){
\lila   \drawline(0,0)(-8,-8)\drawline(0,0)(-16,16)\drawline(-8,-8)(-16,16)
\black} \put(-25,-1){$\mathbb F_0$}\put(-25,7){$\mathbb F_1$}
        \put(-25,15){$\mathbb F_2$}
        \put(4,0){\matrixput(-18,-8)(8,0)3(0,8)4{\circle*2}
        \put(-2,0){\drawline(0,0)(-8,0)\drawline(-8,-8)(-8,8)
                \drawline(-16,16)(-8,0)}}
        \put(-8.7,-12.3){$z_0$}\put(4.5,-1){$z_1$}\put(-15,18){$z_2$}
        \put(-7,10){$w$}

        \put(10,9){$(\blue\l^2\red\m\black z_0\!:\!\blue\l\black z_1\!:\!
                        \blue\l\black z_2\!:\!\red\m\black w)$
                \normalsize$\in(\mathbb C^4\!-\!Z)/(\mathbb C^*)^2$}
        \put(20.5,0){$Z=\{z_0\!=\!w\!=\!0\}\cup\{z_1\!=\!z_2\!=\!0\}$}	\black
\EP								\enddel
\def\ConeSigma{\drawline(0,10)(-20,40)\drawline(0,10)(20,40)
        \dashline2(0,10)(-5,50)\drawline(0,10)(5,30)
        \drawline(5,30)(20,40)\drawline(5,30)(-20,40)
        \drawline(-5,50)(20,40)\drawline(-5,50)(-20,40)}	
				\unitlength=1.2pt	\hspace*{4mm}	%\BC
\BP(100,58)
\put(80,0){\ConeSigma \putlab(-5,52)b{$z_1$}\putlab(22,42)b{$z_3$}
	\putlab(-22,42)b{$z_2$}\putlab(4,32)b{$z_0$}	
	\putlab(0,4)t{\footnotesize(2b)}
	}	    %\put(-32,0){{\bf conifold} {\large(singular geometry)}}}
\put(20,0){\ConeSigma	\putlab(3,32)b{$\b$}\putlab(1,42)b{$\a$}
	\putlab(0,4)t{\footnotesize(2a)}
        \put(0,0){\blue\linethickness{2pt}\drawline(-20,40)(20,40)}\black}
\put(140,0){\ConeSigma	\putlab(9,47)b{$x$}\putlab(-14,46)b{$v$}
%			\putlab(9.9,28)b{$u$}\putlab(-8,28)b{$y$}
	\putlab(0,4)t{\footnotesize(2c)}
			\putlab(9,35)b{$u$}\putlab(-6,28)b{$y$}
        \put(0,0){\blue\linethickness{2pt}\drawline(-5,50)(5,30)}\black}

\put(175,47){$q=(1,1,-1,1)$ \; $\then$ \;
	$(\l z_0:\l z_1:\frac1\l z_2:\frac1\l z_3)$	}
\put(175,34){$x=z_0z_2$, \; $y=z_1z_3$}
\put(175,24){$u=z_1z_2$, \; $v=z_0z_3$}
\put(252,29){$\then$ \; $xy-uv=0$}
\blue	\put(175,12){small resolution: \; \; \; $0\to\IP^1\sim S^2$}
\red	\put(175,1){deformation $xy-uv=\e$: \,$0\to S^3$}
\EP						%\\[8pt]%\hspace*{88mm} \EC
								\del
\stepcounter{figure}						\makeatletter
\immediate\write\@auxout{\string\newlabel{Hirze}{\thefigure}} 	\makeatother
Fig. \thefigure: The Hirzebruch surface $\mathbb F_2$ 
	as blow-up  %$\mathbb F_2$
 	of $W\IP_{211}$.
\HS4.5								
\stepcounter{figure}						\makeatletter
\immediate\write\@auxout{\string\newlabel{ConiF}{\thefigure}} 	\makeatother
Fig. \thefigure: Toric desingularizations (\ref{ConiF}a) and (\ref{ConiF}c)
	of the conifold singularity $xy-uv=0$ in (\ref{ConiF}b).
\end{figure*}							\enddel
\caption{Toric desingularizations (\ref{ConiF}a) and (\ref{ConiF}c)
	of the conifold singularity $xy-uv=0$ in (\ref{ConiF}b).}\label{ConiF}
\vspace*{-9pt}
\end{figure}

The simplest singularity due to a non-simplicial cone is the conifold 
singularity with $v_0+v_1-v_2-v_3=0$ as shown in Fig.\,\ref{ConiF}b. 
The homogeneous coordinates are 
$(\l z_0:\l z_1:\frac1\l z_2:\frac1\l z_3)\in\IC^4/\IC^*$ so that the 
regular $\l$-invariant functions are generated by $x=z_0z_2$, $y=z_1z_3$, 
$u=z_1z_2$, $v=z_0z_3$ and $\IP_\s=\{(x,y,u,v)\in\IC^4~:~xy-uv=0\}$. As
$\s$ is not simplicial, equivalence classes of homogeneous coordinates
are in one-to-one correspondence to conifold points only after some
``bad orbits'' are dropped \cite{CoxHCR,Kreuzer:2006ax}. The toric
resolution of singularities now proceeds by triangulation of the cone.
It amounts to a blow-up $0\to\IP^1$ of the singular point and is called
``small resolution''. The singular
transition between the two different
triangulations is called a flop transition. There is, however, a third
(non-toric) desingularization by deformation of the conifold equation
$xy-uv=\e$, which topologically replaces the singular point by a 3-sphere
\cite{Kreuzer:2006ax}.
The singular transition between a small resolution and a deformation 
is called conifold transition. This will be used in the next section
for constructing new Calabi-Yau manifolds.

%	\subsection{Reflexivity and mirror symmetry}

The dual polytope $\D^\circ$ of a %lattice 
polytope $\D\in M_\IR$ is 
defined by	\hbox{%
$
	\D^\circ=\{y\in N_\IR \,:\,\bra y,x\ket\ge-1\;\forall x\in\D\}.
$}
A lattice polytope $\D$ is called reflexive if $\D^\circ$ also is a lattice 
polytope. 
Let $\N$ be the convex hull of the generators $v_j$ of the rays in
$\S(1)$ for a projective toric variety $\IP_\S$. 
%and $\D_f$ be the Newton polytope of a homogeneous equation. 
It was shown by Batyrev \cite{Bat} that a generic 
member $X_f$ of the hypersurface family with Newton polytope $\D_f$ in $\IP_\S$
is a Calabi-Yau variety if and only if $\D_f^\circ=\N$, so that both polytopes
are reflexive. Moreover, mirror symmetry amounts to the exchange of $\N$
and $\D$. It is manifest at the level of Hodge numbers, for which a
combinatorial formula in terms of numbers of lattice points on faces
of $\D$ and $\N$ can be given \cite{Bat}. If $\S$ is the fan over a maximal
coherent triangulation  of the polytope $\N$ (using all lattice points) then
the singular set of $\IP_\S$ has codimension 4 and is generically avoided by 
the hypersurface equation $f=0$ so that $X_f$ is smooth.

\del
The construction works in any dimension. 
Singularities of maximal crepant projective resolutions (via triangulation
of the polytope $\N$), however, have at least codimension 4 and therefore 
are avoided by generic members of the hypersurface family 
for Calabi-Yau $n$-folds with $n\le3$. For $n>3$ terminal singularities
of the ambient space can prevent the existence of a smooth hypersurface
for a given reflexie polytope.
\enddel

\def\ConifoldHpq{
\h1.25.\h1.28.\h1.29.\h1.30.\h1.31.\h1.32.\h1.33.\h1.34.\h1.35.\h1.36.\h1.37.
        \h1.38.\h1.39.\h1.40.\h1.41.\h1.45.\h1.47.\h1.51.\h1.53.\h1.55.\h1.59.
        \h1.61.\h1.65.\h1.73.\h1.76.\h1.79.\h1.89.\h1.101.\h1.103.\h1.129.
\h2.26.\h2.28.\h2.29.\h2.30.\h2.31.\h2.32.\h2.33.\h2.34.\h2.35.\h2.36.\h2.37.
        \h2.38.\h2.39.\h2.40.\h2.41.\h2.42.\h2.43.\h2.44.\h2.45.\h2.46.\h2.47.
        \h2.48.\h2.49.\h2.50.\h2.51.\h2.52.\h2.53.\h2.54.\h2.55.\h2.56.\h2.57.
        \h2.58.\h2.59.\h2.60.\h2.62.\h2.63.\h2.64.\h2.65.\h2.66.\h2.67.\h2.68.
        \h2.70.\h2.72.\h2.74.\h2.76.\h2.77.\h2.78.\h2.80.\h2.82.\h2.83.\h2.84.
        \h2.86.\h2.88.\h2.90.\h2.96.\h2.100.\h2.102.\h2.112.\h2.116.\h2.128.
\h3.25.\h3.27.\h3.28.\h3.29.\h3.30.\h3.31.\h3.32.\h3.33.\h3.34.\h3.35.\h3.36.
        \h3.37.\h3.38.\h3.39.\h3.40.\h3.41.\h3.42.\h3.43.\h3.44.\h3.45.\h3.46.
        \h3.47.\h3.48.\h3.49.\h3.50.\h3.51.\h3.52.\h3.53.\h3.54.\h3.55.\h3.56.
        \h3.57.\h3.58.\h3.59.\h3.60.\h3.61.\h3.62.\h3.63.\h3.64.\h3.65.\h3.66.
        \h3.67.\h3.68.\h3.69.\h3.70.\h3.71.\h3.72.\h3.73.\h3.75.\h3.76.
        \h3.78.\h3.79.\h3.81.\h3.83.\h3.85.\h3.87.\h3.89.\h3.91.\h3.93.
        \h3.95.\h3.99.\h3.101.\h3.103.\h3.105.\h3.107.\h3.111.\h3.115.
\h4.24.\h4.28.\h4.30.\h4.31.\h4.32.\h4.33.\h4.34.\h4.35.\h4.36.\h4.37.\h4.38.
        \h4.39.\h4.40.\h4.41.\h4.42.\h4.43.\h4.44.\h4.45.\h4.46.\h4.47.\h4.48.
        \h4.49.\h4.50.\h4.51.\h4.52.\h4.53.\h4.54.\h4.55.\h4.56.\h4.57.\h4.58.
        \h4.59.\h4.60.\h4.61.\h4.62.\h4.63.\h4.64.\h4.65.\h4.66.\h4.67.\h4.68.
        \h4.69.\h4.70.\h4.71.\h4.72.\h4.73.\h4.74.\h4.75.\h4.76.\h4.78.\h4.79.
        \h4.80.\h4.81.\h4.82.\h4.84.\h4.86.\h4.88.\h4.89.\h4.90.\h4.91.\h4.92.
        \h4.93.\h4.94.\h4.95.\h4.96.\h4.97.\h4.98.\h4.100.\h4.102.\h4.104.
        \h4.106.\h4.112.
\h5.27.\h5.29.\h5.30.\h5.31.\h5.32.\h5.33.\h5.34.\h5.35.\h5.36.\h5.37.\h5.38.
        \h5.39.\h5.40.\h5.41.\h5.42.\h5.43.\h5.44.\h5.45.\h5.46.\h5.47.\h5.48.
        \h5.49.\h5.50.\h5.51.\h5.52.\h5.53.\h5.54.\h5.55.\h5.56.\h5.57.\h5.58.
        \h5.59.\h5.60.\h5.61.\h5.62.\h5.63.\h5.64.\h5.65.\h5.66.\h5.67.\h5.68.
        \h5.69.\h5.70.\h5.71.\h5.72.\h5.73.\h5.74.\h5.75.\h5.76.\h5.77.\h5.78.
        \h5.79.\h5.80.\h5.81.\h5.82.\h5.83.\h5.85.\h5.86.\h5.87.\h5.88.\h5.89.
        \h5.90.\h5.91.\h5.92.\h5.93.\h5.97.
\h6.28.\h6.30.\h6.31.\h6.32.\h6.34.\h6.35.\h6.36.\h6.37.\h6.38.\h6.39.\h6.40.
        \h6.41.\h6.42.\h6.43.\h6.44.\h6.45.\h6.46.\h6.47.\h6.48.
        \h6.49.\h6.50.\h6.51.\h6.52.\h6.53.\h6.54.\h6.55.\h6.56.\h6.58.
        \h6.59.\h6.60.\h6.61.\h6.62.\h6.63.\h6.64.\h6.65.\h6.66.\h6.67.\h6.68.
        \h6.69.\h6.70.\h6.72.\h6.73.\h6.74.\h6.75.\h6.76.\h6.80.\h6.82.
\h7.27.\h7.29.\h7.30.\h7.31.\h7.33.\h7.34.\h7.35.\h7.37.\h7.38.\h7.39.\h7.40.
        \h7.41.\h7.43.\h7.45.\h7.47.\h7.49.\h7.50.\h7.51.\h7.53.\h7.55.\h7.57.
        \h7.59.\h7.61.\h7.62.\h7.64.\h7.76.
\h8.30.\h8.32.\h8.33.\h8.34.\h8.36.\h8.38.\h8.40.\h8.42.\h8.44.\h8.52.
\h9.31.\h9.33.\h9.37.           \h10.26.\h10.30.\h10.34.\h10.36.
\h11.27.                        \h12.28.                        \h15.23.
}

% \caption{The 15122 (of 30108) hypersurface spectra with $h_{11}\le h_{12}$.
%    The maximal value of $h_{11}+h_{12}$ comes from (251,251) and (491,11).}
\def\HpXXq{					% cutoff at:	min <= 20
						%		max <= 135
\h14.14.\h15.15.\h16.12.\h16.13.\h16.14.\h16.15.\h16.16.\h17.12.
\h17.13.\h17.14.\h17.15.\h17.16.\h17.17.\h18.10.\h18.11.\h18.12.
\h18.13.\h18.14.\h18.15.\h18.16.\h18.17.\h18.18.\h19.7.\h19.9.
\h19.11.\h19.13.\h19.14.\h19.15.\h19.16.\h19.17.\h19.18.\h19.19.
\h20.5.\h20.10.\h20.11.\h20.12.\h20.13.\h20.14.\h20.15.\h20.16.
\h20.17.\h20.18.\h20.19.\h20.20.\h21.1.\h21.9.\h21.11.\h21.12.
\h21.13.\h21.14.\h21.15.\h21.16.\h21.17.\h21.18.\h21.19.\h21.20.
\h22.8.\h22.9.\h22.10.\h22.11.\h22.12.\h22.13.\h22.14.
\h22.15.\h22.16.\h22.17.\h22.18.\h22.19.\h22.20.
\h23.7.\h23.9.\h23.10.\h23.11.\h23.12.\h23.13.\h23.14.\h23.15.
\h23.16.\h23.17.\h23.18.\h23.19.\h23.20.
\h24.8.\h24.9.\h24.10.\h24.11.\h24.12.\h24.13.\h24.14.\h24.15.
\h24.16.\h24.17.\h24.18.\h24.19.\h24.20.
\h25.7.\h25.8.\h25.9.\h25.10.\h25.11.\h25.12.\h25.13.
\h25.14.\h25.15.\h25.16.\h25.17.\h25.18.\h25.19.\h25.20.
\h26.6.\h26.8.\h26.9.\h26.10.
\h26.11.\h26.12.\h26.13.\h26.14.\h26.15.\h26.16.\h26.17.\h26.18.
\h26.19.\h26.20.
\h27.6.\h27.7.\h27.8.\h27.9.\h27.10.\h27.11.\h27.12.\h27.13.
\h27.14.\h27.15.\h27.16.\h27.17.\h27.18.\h27.19.\h27.20.
\h28.4.\h28.6.
\h28.7.\h28.8.\h28.9.\h28.10.\h28.11.\h28.12.\h28.13.\h28.14.
\h28.15.\h28.16.\h28.17.\h28.18.\h28.19.\h28.20.
\h29.2.\h29.5.
\h29.7.\h29.8.\h29.9.\h29.10.\h29.11.\h29.12.\h29.13.\h29.14.
\h29.15.\h29.16.\h29.17.\h29.18.\h29.19.\h29.20.
\h30.6.\h30.7.\h30.8.\h30.9.\h30.10.\h30.11.\h30.12.\h30.13.\h30.14.
\h30.15.\h30.16.\h30.17.\h30.18.\h30.19.\h30.20.
\h31.5.\h31.7.\h31.8.\h31.9.\h31.10.\h31.11.\h31.12.\h31.13.
\h31.14.\h31.15.\h31.16.\h31.17.\h31.18.\h31.19.\h31.20.
\h32.6.\h32.7.\h32.8.\h32.9.\h32.10.\h32.11.
\h32.12.\h32.13.\h32.14.\h32.15.\h32.16.\h32.17.\h32.18.\h32.19.\h32.20.
\h33.5.\h33.6.\h33.7.
\h33.8.\h33.9.\h33.10.\h33.11.\h33.12.\h33.13.\h33.14.\h33.15.
\h33.16.\h33.17.\h33.18.\h33.19.\h33.20.
\h34.4.\h34.6.\h34.7.\h34.8.\h34.9.\h34.10.
\h34.11.\h34.12.\h34.13.\h34.14.\h34.15.\h34.16.\h34.17.\h34.18.
\h34.19.\h34.20.
\h35.5.\h35.7.\h35.8.\h35.9.\h35.10.\h35.11.\h35.12.\h35.13.
\h35.14.\h35.15.\h35.16.\h35.17.\h35.18.\h35.19.\h35.20.
\h36.4.\h36.6.
\h36.7.\h36.8.\h36.9.\h36.10.\h36.11.\h36.12.\h36.13.\h36.14.
\h36.15.\h36.16.\h36.17.\h36.18.\h36.19.\h36.20.
\h37.4.\h37.5.
\h37.6.\h37.7.\h37.8.\h37.9.\h37.10.\h37.11.\h37.12.\h37.13.
\h37.14.\h37.15.\h37.16.\h37.17.\h37.18.\h37.19.\h37.20.
\h38.2.\h38.5.\h38.6.\h38.7.\h38.8.\h38.9.\h38.10.\h38.11.
\h38.12.\h38.13.\h38.14.\h38.15.\h38.16.\h38.17.\h38.18.\h38.19.\h38.20.
\h39.5.\h39.6.\h39.7.\h39.8.\h39.9.
\h39.10.\h39.11.\h39.12.\h39.13.\h39.14.\h39.15.\h39.16.\h39.17.
\h39.18.\h39.19.\h39.20.
\h40.4.\h40.5.
\h40.6.\h40.7.\h40.8.\h40.9.\h40.10.\h40.11.\h40.12.\h40.13.
\h40.14.\h40.15.\h40.16.\h40.17.\h40.18.\h40.19.\h40.20.
\h41.5.\h41.6.\h41.7.\h41.8.\h41.9.
\h41.10.\h41.11.\h41.12.\h41.13.\h41.14.\h41.15.\h41.16.\h41.17.
\h41.18.\h41.19.\h41.20.
\h42.4.\h42.5.\h42.6.\h42.7.\h42.8.\h42.9.\h42.10.\h42.11.
\h42.12.\h42.13.\h42.14.\h42.15.\h42.16.\h42.17.\h42.18.\h42.19.
\h42.20.
\h43.3.\h43.5.\h43.6.\h43.7.\h43.8.\h43.9.\h43.10.\h43.11.\h43.12.
\h43.13.\h43.14.\h43.15.\h43.16.\h43.17.\h43.18.\h43.19.\h43.20.
\h44.4.\h44.5.\h44.6.\h44.7.\h44.8.\h44.9.\h44.10.\h44.11.\h44.12.
\h44.13.\h44.14.\h44.15.\h44.16.\h44.17.\h44.18.\h44.19.\h44.20.
\h45.3.\h45.5.\h45.6.\h45.7.\h45.8.\h45.9.\h45.10.\h45.11.
\h45.12.\h45.13.\h45.14.\h45.15.\h45.16.\h45.17.\h45.18.\h45.19.
\h45.20.
\h46.4.\h46.5.\h46.6.\h46.7.\h46.8.\h46.9.
\h46.10.\h46.11.\h46.12.\h46.13.\h46.14.\h46.15.\h46.16.\h46.17.
\h46.18.\h46.19.\h46.20.
\h46.21.\h46.22.\h46.23.\h46.24.\h46.25.
\h46.26.\h46.27.\h46.28.\h46.29.\h46.30.\h46.31.\h46.32.\h46.33.
\h46.34.\h46.35.\h46.36.\h46.37.\h46.38.\h46.39.\h46.40.\h46.41.
\h46.42.\h46.43.\h46.44.\h46.45.\h46.46.\h47.5.\h47.6.\h47.7.
\h47.8.\h47.9.\h47.10.\h47.11.\h47.12.\h47.13.\h47.14.\h47.15.
\h47.16.\h47.17.\h47.18.\h47.19.\h47.20.\h47.21.\h47.22.\h47.23.
\h47.24.\h47.25.\h47.26.\h47.27.\h47.28.\h47.29.\h47.30.\h47.31.
\h47.32.\h47.33.\h47.34.\h47.35.\h47.36.\h47.37.\h47.38.\h47.39.
\h47.40.\h47.41.\h47.42.\h47.43.\h47.44.\h47.45.\h47.46.\h47.47.
\h48.4.\h48.5.\h48.6.\h48.7.\h48.8.\h48.9.\h48.10.\h48.11.
\h48.12.\h48.13.\h48.14.\h48.15.\h48.16.\h48.17.\h48.18.\h48.19.
\h48.20.
\h49.4.\h49.5.\h49.6.
\h49.7.\h49.8.\h49.9.\h49.10.\h49.11.\h49.12.\h49.13.\h49.14.
\h49.15.\h49.16.\h49.17.\h49.18.\h49.19.\h49.20.
\h50.4.\h50.5.\h50.6.\h50.7.\h50.8.
\h50.9.\h50.10.\h50.11.\h50.12.\h50.13.\h50.14.\h50.15.\h50.16.
\h50.17.\h50.18.\h50.19.\h50.20.
\h51.3.\h51.5.\h51.6.\h51.7.\h51.8.\h51.9.
\h51.10.\h51.11.\h51.12.\h51.13.\h51.14.\h51.15.\h51.16.\h51.17.
\h51.18.\h51.19.\h51.20.
\h52.4.\h52.5.\h52.6.\h52.7.\h52.8.\h52.9.
\h52.10.\h52.11.\h52.12.\h52.13.\h52.14.\h52.15.\h52.16.\h52.17.
\h52.18.\h52.19.\h52.20.
\h53.5.\h53.6.\h53.7.\h53.8.\h53.9.
\h53.10.\h53.11.\h53.12.\h53.13.\h53.14.\h53.15.\h53.16.\h53.17.
\h53.18.\h53.19.\h53.20.
\h54.4.\h54.5.\h54.6.\h54.7.
\h54.8.\h54.9.\h54.10.\h54.11.\h54.12.\h54.13.\h54.14.\h54.15.
\h54.16.\h54.17.\h54.18.\h54.19.\h54.20.
\h55.4.
\h55.5.\h55.6.\h55.7.\h55.8.\h55.9.\h55.10.\h55.11.\h55.12.
\h55.13.\h55.14.\h55.15.\h55.16.\h55.17.\h55.18.\h55.19.\h55.20.
\h56.4.\h56.5.\h56.6.\h56.7.\h56.8.
\h56.9.\h56.10.\h56.11.\h56.12.\h56.13.\h56.14.\h56.15.\h56.16.
\h56.17.\h56.18.\h56.19.\h56.20.
\h57.3.\h57.4.\h57.5.\h57.6.\h57.7.\h57.8.\h57.9.\h57.10.
\h57.11.\h57.12.\h57.13.\h57.14.\h57.15.\h57.16.\h57.17.\h57.18.
\h57.19.\h57.20.
\h58.4.
\h58.5.\h58.6.\h58.7.\h58.8.\h58.9.\h58.10.\h58.11.\h58.12.
\h58.13.\h58.14.\h58.15.\h58.16.\h58.17.\h58.18.\h58.19.\h58.20.
\h59.3.\h59.4.
\h59.5.\h59.6.\h59.7.\h59.8.\h59.9.\h59.10.\h59.11.\h59.12.
\h59.13.\h59.14.\h59.15.\h59.16.\h59.17.\h59.18.\h59.19.\h59.20.
\h60.4.
\h60.5.\h60.6.\h60.7.\h60.8.\h60.9.\h60.10.\h60.11.\h60.12.
\h60.13.\h60.14.\h60.15.\h60.16.\h60.17.\h60.18.\h60.19.\h60.20.
\h61.4.\h61.5.\h61.6.\h61.7.\h61.8.\h61.9.\h61.10.\h61.11.
\h61.12.\h61.13.\h61.14.\h61.15.\h61.16.\h61.17.\h61.18.\h61.19.
\h61.20.
\h62.4.\h62.5.\h62.6.\h62.7.\h62.8.\h62.9.
\h62.10.\h62.11.\h62.12.\h62.13.\h62.14.\h62.15.\h62.16.\h62.17.
\h62.18.\h62.19.\h62.20.
\h63.3.\h63.4.\h63.5.
\h63.6.\h63.7.\h63.8.\h63.9.\h63.10.\h63.11.\h63.12.\h63.13.
\h63.14.\h63.15.\h63.16.\h63.17.\h63.18.\h63.19.\h63.20.
\h64.4.\h64.5.\h64.6.\h64.7.\h64.8.\h64.9.
\h64.10.\h64.11.\h64.12.\h64.13.\h64.14.\h64.15.\h64.16.\h64.17.
\h64.18.\h64.19.\h64.20.
\h65.3.
\h65.4.\h65.5.\h65.6.\h65.7.\h65.8.\h65.9.\h65.10.\h65.11.
\h65.12.\h65.13.\h65.14.\h65.15.\h65.16.\h65.17.\h65.18.\h65.19.
\h65.20.
\h66.3.\h66.4.
\h66.5.\h66.6.\h66.7.\h66.8.\h66.9.\h66.10.\h66.11.\h66.12.
\h66.13.\h66.14.\h66.15.\h66.16.\h66.17.\h66.18.\h66.19.\h66.20.
\h67.3.\h67.4.
\h67.5.\h67.6.\h67.7.\h67.8.\h67.9.\h67.10.\h67.11.\h67.12.
\h67.13.\h67.14.\h67.15.\h67.16.\h67.17.\h67.18.\h67.19.\h67.20.
\h68.4.
\h68.5.\h68.6.\h68.7.\h68.8.\h68.9.\h68.10.\h68.11.\h68.12.
\h68.13.\h68.14.\h68.15.\h68.16.\h68.17.\h68.18.\h68.19.\h68.20.
\h69.3.\h69.4.\h69.5.\h69.6.\h69.7.\h69.8.\h69.9.\h69.10.
\h69.11.\h69.12.\h69.13.\h69.14.\h69.15.\h69.16.\h69.17.\h69.18.
\h69.19.\h69.20.
\h70.4.\h70.5.\h70.6.\h70.7.\h70.8.
\h70.9.\h70.10.\h70.11.\h70.12.\h70.13.\h70.14.\h70.15.\h70.16.
\h70.17.\h70.18.\h70.19.\h70.20.
\h71.3.\h71.4.
\h71.5.\h71.6.\h71.7.\h71.8.\h71.9.\h71.10.\h71.11.\h71.12.
\h71.13.\h71.14.\h71.15.\h71.16.\h71.17.\h71.18.\h71.19.\h71.20.
\h72.3.\h72.4.\h72.5.\h72.6.\h72.7.
\h72.8.\h72.9.\h72.10.\h72.11.\h72.12.\h72.13.\h72.14.\h72.15.
\h72.16.\h72.17.\h72.18.\h72.19.\h72.20.
\h73.3.\h73.4.\h73.5.\h73.6.\h73.7.\h73.8.\h73.9.
\h73.10.\h73.11.\h73.12.\h73.13.\h73.14.\h73.15.\h73.16.\h73.17.
\h73.18.\h73.19.\h73.20.
\h74.2.\h74.4.\h74.5.\h74.6.\h74.7.\h74.8.\h74.9.\h74.10.
\h74.11.\h74.12.\h74.13.\h74.14.\h74.15.\h74.16.\h74.17.\h74.18.
\h74.19.\h74.20.
\h75.3.\h75.5.\h75.6.\h75.7.\h75.8.\h75.9.\h75.10.\h75.11.
\h75.12.\h75.13.\h75.14.\h75.15.\h75.16.\h75.17.\h75.18.\h75.19.
\h75.20.
\h76.3.\h76.4.\h76.5.\h76.6.\h76.7.\h76.8.\h76.9.\h76.10.
\h76.11.\h76.12.\h76.13.\h76.14.\h76.15.\h76.16.\h76.17.\h76.18.
\h76.19.\h76.20.
\h77.3.\h77.4.\h77.5.\h77.6.\h77.7.\h77.8.
\h77.9.\h77.10.\h77.11.\h77.12.\h77.13.\h77.14.\h77.15.\h77.16.
\h77.17.\h77.18.\h77.19.\h77.20.
\h78.3.\h78.4.\h78.5.
\h78.6.\h78.7.\h78.8.\h78.9.\h78.10.\h78.11.\h78.12.\h78.13.
\h78.14.\h78.15.\h78.16.\h78.17.\h78.18.\h78.19.\h78.20.
\h79.3.\h79.4.\h79.5.\h79.6.\h79.7.\h79.8.\h79.9.
\h79.10.\h79.11.\h79.12.\h79.13.\h79.14.\h79.15.\h79.16.\h79.17.
\h79.18.\h79.19.\h79.20.
\h80.4.\h80.5.
\h80.6.\h80.7.\h80.8.\h80.9.\h80.10.\h80.11.\h80.12.\h80.13.
\h80.14.\h80.15.\h80.16.\h80.17.\h80.18.\h80.19.\h80.20.
\h81.3.\h81.4.\h81.5.\h81.6.\h81.7.
\h81.8.\h81.9.\h81.10.\h81.11.\h81.12.\h81.13.\h81.14.\h81.15.
\h81.16.\h81.17.\h81.18.\h81.19.\h81.20.
\h82.4.\h82.5.\h82.6.\h82.7.\h82.8.\h82.9.
\h82.10.\h82.11.\h82.12.\h82.13.\h82.14.\h82.15.\h82.16.\h82.17.
\h82.18.\h82.19.\h82.20.
\h83.2.\h83.3.\h83.5.\h83.6.\h83.7.\h83.8.\h83.9.
\h83.10.\h83.11.\h83.12.\h83.13.\h83.14.\h83.15.\h83.16.\h83.17.
\h83.18.\h83.19.\h83.20.
\h84.2.\h84.3.\h84.4.\h84.5.\h84.6.\h84.7.
\h84.8.\h84.9.\h84.10.\h84.11.\h84.12.\h84.13.\h84.14.\h84.15.
\h84.16.\h84.17.\h84.18.\h84.19.\h84.20.
\h85.3.\h85.4.\h85.5.
\h85.6.\h85.7.\h85.8.\h85.9.\h85.10.\h85.11.\h85.12.\h85.13.
\h85.14.\h85.15.\h85.16.\h85.17.\h85.18.\h85.19.\h85.20.
\h86.2.\h86.4.\h86.5.\h86.6.\h86.7.\h86.8.\h86.9.\h86.10.
\h86.11.\h86.12.\h86.13.\h86.14.\h86.15.\h86.16.\h86.17.\h86.18.
\h86.19.\h86.20.
\h87.3.\h87.5.\h87.6.\h87.7.
\h87.8.\h87.9.\h87.10.\h87.11.\h87.12.\h87.13.\h87.14.\h87.15.
\h87.16.\h87.17.\h87.18.\h87.19.\h87.20.
\h88.4.\h88.5.\h88.6.\h88.7.\h88.8.\h88.9.\h88.10.\h88.11.
\h88.12.\h88.13.\h88.14.\h88.15.\h88.16.\h88.17.\h88.18.\h88.19.
\h88.20.
\h89.3.\h89.4.\h89.5.
\h89.6.\h89.7.\h89.8.\h89.9.\h89.10.\h89.11.\h89.12.\h89.13.
\h89.14.\h89.15.\h89.16.\h89.17.\h89.18.\h89.19.\h89.20.
\h90.2.\h90.4.\h90.5.\h90.6.\h90.7.
\h90.8.\h90.9.\h90.10.\h90.11.\h90.12.\h90.13.\h90.14.\h90.15.
\h90.16.\h90.17.\h90.18.\h90.19.\h90.20.
\h91.3.\h91.4.\h91.5.\h91.6.\h91.7.
\h91.8.\h91.9.\h91.10.\h91.11.\h91.12.\h91.13.\h91.14.\h91.15.
\h91.16.\h91.17.\h91.18.\h91.19.\h91.20.
\h92.2.\h92.4.\h92.5.\h92.6.\h92.7.
\h92.8.\h92.9.\h92.10.\h92.11.\h92.12.\h92.13.\h92.14.\h92.15.
\h92.16.\h92.17.\h92.18.\h92.19.\h92.20.
\h93.3.\h93.4.\h93.5.\h93.6.
\h93.7.\h93.8.\h93.9.\h93.10.\h93.11.\h93.12.\h93.13.\h93.14.
\h93.15.\h93.16.\h93.17.\h93.18.\h93.19.\h93.20.
\h94.4.\h94.5.
\h94.6.\h94.7.\h94.8.\h94.9.\h94.10.\h94.11.\h94.12.\h94.13.
\h94.14.\h94.15.\h94.16.\h94.17.\h94.18.\h94.19.\h94.20.
\h95.2.\h95.3.\h95.5.\h95.6.\h95.7.\h95.8.\h95.9.\h95.10.
\h95.11.\h95.12.\h95.13.\h95.14.\h95.15.\h95.16.\h95.17.\h95.18.
\h95.19.\h95.20.
\h96.4.\h96.6.\h96.7.
\h96.8.\h96.9.\h96.10.\h96.11.\h96.12.\h96.13.\h96.14.\h96.15.
\h96.16.\h96.17.\h96.18.\h96.19.\h96.20.
\h97.4.
\h97.5.\h97.6.\h97.7.\h97.8.\h97.9.\h97.10.\h97.11.\h97.12.
\h97.13.\h97.14.\h97.15.\h97.16.\h97.17.\h97.18.\h97.19.\h97.20.
\h98.4.\h98.5.\h98.6.\h98.7.
\h98.8.\h98.9.\h98.10.\h98.11.\h98.12.\h98.13.\h98.14.\h98.15.
\h98.16.\h98.17.\h98.18.\h98.19.\h98.20.
\h99.3.\h99.5.\h99.6.\h99.7.\h99.8.\h99.9.
\h99.10.\h99.11.\h99.12.\h99.13.\h99.14.\h99.15.\h99.16.\h99.17.
\h99.18.\h99.19.\h99.20.
\h100.4.\h100.6.\h100.7.\h100.8.\h100.9.\h100.10.\h100.11.
\h100.12.\h100.13.\h100.14.\h100.15.\h100.16.\h100.17.\h100.18.\h100.19.
\h100.20.
\h101.1.
\h101.4.\h101.5.\h101.6.\h101.7.\h101.8.\h101.9.\h101.10.\h101.11.
\h101.12.\h101.13.\h101.14.\h101.15.\h101.16.\h101.17.\h101.18.\h101.19.
\h101.20.
\h102.2.\h102.4.\h102.6.\h102.7.\h102.8.\h102.9.\h102.10.
\h102.11.\h102.12.\h102.13.\h102.14.\h102.15.\h102.16.\h102.17.\h102.18.
\h102.19.\h102.20.
\h103.1.\h103.3.\h103.5.\h103.6.\h103.7.\h103.8.
\h103.9.\h103.10.\h103.11.\h103.12.\h103.13.\h103.14.\h103.15.\h103.16.
\h103.17.\h103.18.\h103.19.\h103.20.
\h104.4.\h104.5.
\h104.6.\h104.7.\h104.8.\h104.9.\h104.10.\h104.11.\h104.12.\h104.13.
\h104.14.\h104.15.\h104.16.\h104.17.\h104.18.\h104.19.\h104.20.
\h105.3.\h105.5.\h105.6.\h105.7.\h105.8.\h105.9.\h105.10.
\h105.11.\h105.12.\h105.13.\h105.14.\h105.15.\h105.16.\h105.17.\h105.18.
\h105.19.\h105.20.
\h106.2.\h106.4.\h106.6.\h106.7.\h106.8.\h106.9.\h106.10.
\h106.11.\h106.12.\h106.13.\h106.14.\h106.15.\h106.16.\h106.17.\h106.18.
\h106.19.\h106.20.
\h107.3.\h107.5.\h107.6.\h107.7.\h107.8.\h107.9.\h107.10.
\h107.11.\h107.12.\h107.13.\h107.14.\h107.15.\h107.16.\h107.17.\h107.18.
\h107.19.\h107.20.
\h108.4.\h108.5.\h108.6.\h108.7.
\h108.8.\h108.9.\h108.10.\h108.11.\h108.12.\h108.13.\h108.14.\h108.15.
\h108.16.\h108.17.\h108.18.\h108.19.\h108.20.
\h109.4.
\h109.5.\h109.6.\h109.7.\h109.8.\h109.9.\h109.10.\h109.11.\h109.12.
\h109.13.\h109.14.\h109.15.\h109.16.\h109.17.\h109.18.\h109.19.\h109.20.
\h110.4.\h110.5.\h110.6.
\h110.7.\h110.8.\h110.9.\h110.10.\h110.11.\h110.12.\h110.13.\h110.14.
\h110.15.\h110.16.\h110.17.\h110.18.\h110.19.\h110.20.
\h111.3.\h111.5.\h111.6.
\h111.7.\h111.8.\h111.9.\h111.10.\h111.11.\h111.12.\h111.13.\h111.14.
\h111.15.\h111.16.\h111.17.\h111.18.\h111.19.\h111.20.
\h112.4.\h112.6.\h112.7.\h112.8.
\h112.9.\h112.10.\h112.11.\h112.12.\h112.13.\h112.14.\h112.15.\h112.16.
\h112.17.\h112.18.\h112.19.\h112.20.
\h113.5.
\h113.7.\h113.8.\h113.9.\h113.10.\h113.11.\h113.12.\h113.13.\h113.14.
\h113.15.\h113.16.\h113.17.\h113.18.\h113.19.\h113.20.
\h114.4.\h114.6.\h114.7.\h114.8.\h114.9.\h114.10.
\h114.11.\h114.12.\h114.13.\h114.14.\h114.15.\h114.16.\h114.17.\h114.18.
\h114.19.\h114.20.
\h115.3.
\h115.5.\h115.6.\h115.7.\h115.8.\h115.9.\h115.10.\h115.11.\h115.12.
\h115.13.\h115.14.\h115.15.\h115.16.\h115.17.\h115.18.\h115.19.\h115.20.
\h116.2.\h116.4.\h116.5.
\h116.6.\h116.7.\h116.8.\h116.9.\h116.10.\h116.11.\h116.12.\h116.13.
\h116.14.\h116.15.\h116.16.\h116.17.\h116.18.\h116.19.\h116.20.
\h117.5.\h117.6.\h117.7.
\h117.9.\h117.10.\h117.11.\h117.12.\h117.13.\h117.14.\h117.15.\h117.16.
\h117.17.\h117.18.\h117.19.\h117.20.
\h118.4.\h118.6.\h118.8.
\h118.9.\h118.10.\h118.11.\h118.12.\h118.13.\h118.14.\h118.15.\h118.16.
\h118.17.\h118.18.\h118.19.\h118.20.
\h119.3.\h119.5.\h119.7.\h119.8.\h119.9.\h119.10.\h119.11.\h119.12.
\h119.13.\h119.14.\h119.15.\h119.16.\h119.17.\h119.18.\h119.19.\h119.20.
\h120.2.\h120.4.\h120.6.\h120.7.\h120.8.\h120.9.
\h120.10.\h120.11.\h120.12.\h120.13.\h120.14.\h120.15.\h120.16.\h120.17.
\h120.18.\h120.19.\h120.20.
\h121.4.\h121.5.\h121.7.\h121.8.\h121.9.\h121.10.
\h121.11.\h121.12.\h121.13.\h121.14.\h121.15.\h121.16.\h121.17.\h121.18.
\h121.19.\h121.20.
\h122.2.\h122.4.\h122.5.
\h122.6.\h122.7.\h122.8.\h122.9.\h122.10.\h122.11.\h122.12.\h122.13.
\h122.14.\h122.15.\h122.16.\h122.17.\h122.18.\h122.19.\h122.20.
\h123.3.\h123.5.\h123.7.\h123.8.\h123.9.\h123.10.
\h123.11.\h123.12.\h123.13.\h123.14.\h123.15.\h123.16.\h123.17.\h123.18.
\h123.19.\h123.20.
\h124.4.\h124.6.\h124.7.\h124.8.
\h124.9.\h124.10.\h124.11.\h124.12.\h124.13.\h124.14.\h124.15.\h124.16.
\h124.17.\h124.18.\h124.19.\h124.20.
\h125.5.\h125.7.\h125.8.\h125.9.\h125.10.
\h125.11.\h125.12.\h125.13.\h125.14.\h125.15.\h125.16.\h125.17.\h125.18.
\h125.19.\h125.20.
\h126.4.
\h126.6.\h126.8.\h126.9.\h126.10.\h126.11.\h126.12.\h126.13.\h126.14.
\h126.15.\h126.16.\h126.17.\h126.18.\h126.19.\h126.20.
\h127.3.\h127.5.\h127.6.\h127.7.
\h127.9.\h127.10.\h127.11.\h127.12.\h127.13.\h127.14.\h127.15.\h127.16.
\h127.17.\h127.18.\h127.19.\h127.20.
\h128.2.
\h128.4.\h128.5.\h128.6.\h128.8.\h128.9.\h128.10.\h128.11.\h128.12.
\h128.13.\h128.14.\h128.15.\h128.16.\h128.17.\h128.18.\h128.19.\h128.20.
\h129.5.\h129.9.\h129.10.\h129.11.\h129.12.\h129.13.\h129.14.\h129.15.
\h129.16.\h129.17.\h129.18.\h129.19.\h129.20.
\h130.4.\h130.7.\h130.8.\h130.9.\h130.10.\h130.11.\h130.12.\h130.13.
\h130.14.\h130.15.\h130.16.\h130.17.\h130.18.\h130.19.\h130.20.
\h131.3.\h131.5.\h131.7.\h131.8.\h131.9.\h131.10.\h131.11.
\h131.12.\h131.13.\h131.14.\h131.15.\h131.16.\h131.17.\h131.18.\h131.19.
\h131.20.
\h132.2.\h132.6.\h132.8.
\h132.9.\h132.10.\h132.11.\h132.12.\h132.13.\h132.14.\h132.15.\h132.16.
\h132.17.\h132.18.\h132.19.\h132.20.
\h133.5.\h133.7.\h133.9.\h133.10.\h133.11.\h133.12.
\h133.13.\h133.14.\h133.15.\h133.16.\h133.17.\h133.18.\h133.19.\h133.20.
\h134.8.
\h134.9.\h134.10.\h134.11.\h134.12.\h134.13.\h134.14.\h134.15.\h134.16.
\h134.17.\h134.18.\h134.19.\h134.20.
\h135.5.
\h135.7.\h135.9.\h135.10.\h135.11.\h135.12.\h135.13.\h135.14.\h135.15.
\h135.16.\h135.17.\h135.18.\h135.19.\h135.20.
}						%% cut off a p<=20 q <=135

\section{Conifold transitions and mirror symmetry}

In physics conifold transitions were first studied by Candelas et al., who
noted that the singularity is located at finite distance in moduli space 
\cite{Rolling}. The physics of the resulting CFT singularity was later 
understood by Strominger \cite{Strominger:1995qi} in terms of wrapped 
D-branes that become massless black holes in 4d at the transition point.
In mathematics, toric degenerations of Calabi-Yau hypersurfaces in
Grassmannians \cite{Grass} and Flag manifolds \cite{Flag} 
were used for mirror constructions. %\cite{Grass,Flag}. 
Our construction \cite{ccy} is a generalization of this idea.
\VS-7

\begin{table}[h]
  \caption{Numbers of polytopes and Hodge data for hypersurfaces (H) and
	 conifold CYs (C).}	%% \begin{tabular}{@{}lll@{}} \hline
  \begin{tabular}{cccl\TVR{3.5}{.5}} \hline\VR{3.6}{1.9}
  $h_{11}$ & \#$(\D)_H$ &\#$(\D)_C$ & $(h_{12})_C$ \\\hline
% 8871 CYs with h12 = 21,23-51,53,55,59,61,65,73,76,79,89,101,103,129 smooth: 
1 & 5& 210 & 25,28-41,45,47,51,53,55,59,61,65,73,76,79,89,101,103,129\\
% 43080 CYs with h12 = 22,24-80,82-90,96,100,102,103,111,112,116,128 smooth:
2 & 36& 3470 & 
    26,28-60,62-68,70,72,74,76,77,78,80,82-84,86,88,90,96,100,102,112,116,128\\
3&244&11389&25,27-73,75-79,81,83,85,87,89,91,93,95,99,101,103,105,107,111,115\\
4 & 1197 & 10264  & 24,28,30-76,78-82,84,86,88-98,100,102,104,106,112\\
5 & 4990 & 3898 & 27,29,30-83,85-93,97\\
6 & 17101&  815 & 28,30-32,34-56,58-70,72-76,80,82\\
%	7 & 140 &27,29-31,33-35,37-41,43,45,47,49-51,53,55,57,59,61,62,64,76\\
	\hline	\end{tabular}
  \label{tab:ccy}
\end{table}

% In \cite{ccy} we started from the list of Calabi-Yau hypersurfaces in toric
% ambient spaces \cite{c4d,wwwCY} and studied conifold transitions 
% $198849$	Namikawa we found $30241$ 

\begin{figure}
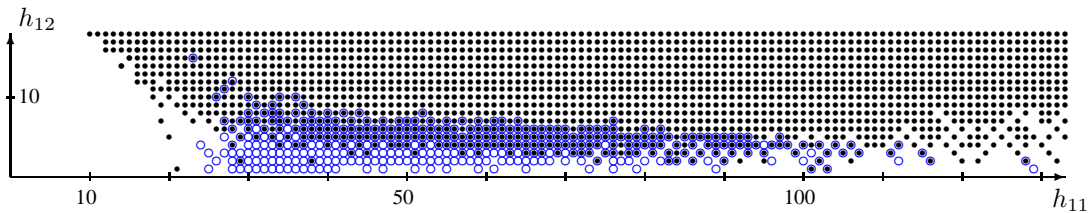

\BC     \def\MaxCHC{18}   \def\Msum{133}        \def\CSH{.55} \def\CSC{.9} 
\unitlength=3pt                
\BP(80,21)(28,1)        \putvec(0,0,1,0,133)    \putvec(0,0,0,1,18)
        \put(1,\MaxCHC){\put(0,1){$h_{12}$}}  \put(\Msum,-3.8){$\!\!\!h_{11}$}
        \put(1,9.3){\fns10}\put(8.2,-3.5){\fns10}\put(48.2,-3.5){\fns50}
        \put(97.5,-3.5){\fns100}
        \def\Xlab#1 {\put(#1,0){\drawline(0,-.5)(0,.5)}}
        \def\Ylab#1 {\put(0,#1){\drawline(.5,0)(-.5,0)}}
        \Xlab10 \Xlab20 \Xlab30 \Xlab40 \Xlab50 \Xlab60 \Xlab70 \Xlab80 
        \Xlab90 \Xlab100 \Xlab110 \Xlab120 \Xlab130 %\Xlab140 \Xlab150   
        \Ylab10 %\Ylab20
\def\h#1.#2.{ \ifnum \MaxCHC < #1 \else \put(#2,#1){\blue \circle{\CSC}} \fi }
{
        \ConifoldHpq    %\FanoVdHpq             
}
\def\h#1.#2.{ \ifnum \MaxCHC < #1 \else \ifnum #2 > \Msum \else
                        \put(#2,#1){\black\circle*{\CSH}} \fi \fi
              \ifnum \MaxCHC < #2 \else \ifnum #1 > \Msum \else
                        \put(#1,#2){\black\circle*{\CSH}} \fi \fi}
{
        \HpXXq		%\INPUT {inc/hyperHpq}   
}
\EP\EC
\caption{Hodge data of Conifold Calabi-Yau manifolds (circles) in the
	background of hypersurface data (dots).}\label{fig:ccy}\VS-12
\end{figure}

Since we want a combinatorial description of the conifold singularities we
consider 4-dimensional reflexive polytopes $\D^\circ$ whose 2-faces all are 
either basic triangles or parallelograms $\th^\circ$ of minimal volume.
There are $198\,849$ polytopes with this property.
% (whose triangulations yield two basic triangles). 
For toric varieties the dimension of a singular set is equal to the 
codimension of the corresponding singular cone. 
We hence obtain a curve of conifold singularities that 
intersects with a generic hypersurface $X_f$ in $k_\th$ points, 
where $k_\th$ is the length (i.e. number of lattic points $-1$) of the edge 
$\th$ of $\D$ that is dual to $\th^\circ$. The total number of conifold points
of $X_f$ is $k=\sum k_\th$. Differently from the usual procedure of choosing
a maximal crepant projective resolution, which would imply a triangulation of
all $\th^\circ$, we keep the singular conifold curves in the ambient space 
and try to obtain a smooth CY by deformation. Namikawa found a criterion for
such a deformation to exist \cite{Smoothing} which can be translated into
combinatorics \cite{ccy} and leaves us with 30241 smoothable cases. Since
the conifold transition blows down a number of $\IP^1$'s and replaces them
by $S^3$'s the Picard number is reduced and we get a sizable number of 
new CYs with small $h_{11}$, as listed in Table \ref{tab:ccy} and 
shown in Fig. \ref{fig:ccy}. 
The computations have been performed with PALP \cite{PALP,wwwCY}. 
The complete data can be found on the internet \cite{wwwCCY}.

So far only the 210 polytopes leading to $h_{11}=1$ have been analyzed in 
some detail. Computing the tripple intersection and the second Chern class
we find 68 diffeomorphism types. Our proposed mirror construction uses the
symplectic surgery condition of Smith, Thomas and Yau \cite{STY} and amounts
to a specialization of the complex structure moduli \cite{ccy}. We thus
computed 30 different Picard-Fuchs operators.

\del
\begin{table}[h]
  \caption{Hodge numbers and numbers of polytopes of conifold Calabi-Yau 
	manifolds}	%% \begin{tabular}{@{}lll@{}} \hline
  \begin{tabular}{ccl\TVR{3.5}{.5}} \hline\VR{3.6}{1.9}
  $h_{11}$ & \#$(\D)_C$ & $h_{12}$ \\\hline
% 8871 CYs with h12 = 21,23-51,53,55,59,61,65,73,76,79,89,101,103,129 smooth: 
	1 & 210 & 25,28-41,45,47,51,53,55,59,61,65,73,76,79,89,101,103,129\\
% 43080 CYs with h12 = 22,24-80,82-90,96,100,102,103,111,112,116,128 smooth:
	2 & 3470 & 
26,28-60,62-68,70,72,74,76,77,78,80,82-84,86,88,90,96,100,102,112,116,128\\
3& 11389 &25,27-73,75-79,81,83,85,87,89,91,93,95,99,101,103,105,107,111,115\\
	4 & 10264  & 24,28,30-76,78-82,84,86,88-98,100,102,104,106,112\\
	5 & 3898 & 27,29,30-83,85-93,97\\
	6 & 815 & 28,30-32,34-56,58-70,72-76,80,82\\
%	7 & 140 &27,29-31,33-35,37-41,43,45,47,49-51,53,55,57,59,61,62,64,76\\
	\hline	\end{tabular}
  \label{tab:h12}
\end{table}

\begin{table}[h]
  \caption{Polytopes $\D$, Euler numbers, and diffeomorphism types
	for hypersurface (H) and conifold (C) CYs.}
  \begin{tabular}{crrccc\TVR{3.5}{.5}} \hline\VR{3.6}{1.9}
$h_{11}$ & $\#(\D)_C$ &$\#(\D)_H$& $\#(Euler)_C$&$\#(Euler)_H$
	&\#(diffeo. types)$_C$\\\hline
1 & \blue210\black &5 & \blue30\black & 5&\blue68\black\\
2 & \blue3470\black&36 & \blue60\black &18& \blue?\black\\
3 & 11389 &244 & 68 &42& \blue?\black\\
4 & 10264 &1197 & 72 & 87&\blue?\black\\
5 & 3808 &4990 & 66 & 113&\blue?\black\\
6 & 815 &17101 & 47 & 128&\blue?\black\\
% 7 & 140 &50376 & 26 & 149&\blue?\black\\
% 8 & 35 &128165 & 10 & 158&\blue?\black\\
	\hline  \end{tabular}
\end{table}
\enddel

\section{Directions}

For the conifold CYs discussed in section 3 much remains to be
done, as they have been studied in some detail only for $h_{11}=1$.
Even for this case the Picard-Fuchs operators are still unknown for all CYs
with $h_{12}\le36$ and there are indications that in some cases our mirror
proposal does not work.%				% \cite{Pfaff}
\footnote{G. Almkvist and D.~van Straten (private communication).}
It would be interesting to enumerate the diffeomorphism types (including
torsion in cohomology \cite{IC}) of all known
examples for small Picard numbers, say $h_{11}\le5$, so that equivalences
and the structure of the web can be analyzed.
Moreover, other types of singular transitions should be investigated.

Many interesting geometries cannot be realized with hypersurfaces
\cite{kkrs,WSI}. While many examples of complete intersections 
are known \cite{wpci,kkrs}, a complete enumeration, at least for small
codimension and Picard number, would be important.
This should be feasible \cite{Kreuzer:2006ax}
via an enumeration of reflexive Gorenstein cones 
\cite{BN}.
% \footnote{This was noted independently by Borisov (private communication).}
% using Skarke's algorithm \cite{Skarke:1996hq}, should be possible.
%

In a recent study of free permutation quotients
Candelas et al. \cite{tip} %\cite{Triado,Candelas:2008wb} 
succeeded in populating the
realm where both Hodge numbers are small. Clearly combinations of these
tools and the study of connections to other constructions would be of 
interest in order to get a better understanding of the web as well as
for phenomenological purposes.

%%%	\newpage

\end{document}